# Stability of Complex Biomolecular Structures: Van der Waals, Hydrogen Bond Cooperativity, and Nuclear Quantum Effects

Mariana Rossi,[*,†,¶] Wei Fang,[‡] and Angelos Michaelides[‡]

*Physical and Theoretical Chemistry Lab, University of Oxford, South Parks Road, OX1 3QZ Oxford, UK, and Thomas Young Centre, London Centre for Nanotechnology, and Department of Chemistry, University College London, 17-19 Gordon Street, WC1H 0AH London, UK*

E-mail: mariana.rossi@chem.ox.ac.uk

---

[*]To whom correspondence should be addressed
[†]Physical and Theoretical Chemistry Lab, University of Oxford, South Parks Road, OX1 3QZ Oxford, UK
[‡]Thomas Young Centre, London Centre for Nanotechnology, and Department of Chemistry, University College London, 17-19 Gordon Street, WC1H 0AH London, UK
[¶]St. Edmund Hall, Queen's Lane, OX1 4AR Oxford, UK

**Abstract**

Biomolecules are complex systems stabilised by a delicate balance of weak interactions, making it important to assess all energetic contributions in an accurate manner. However, it is *a priori* unclear which contributions make more of an impact. Here we examine stacked polyglutamine (polyQ) strands, a peptide repeat often found in amyloid aggregates. We investigate the role of hydrogen bond (HB) cooperativity, van der Waals (vdW) dispersion interactions, and quantum contributions to free energies, including anharmonicities through density-functional theory and *ab initio* path integral simulations. Of these various factors, we find that the largest impact on structural stabilisation comes from vdW interactions. HB cooperativity is the second largest contribution, as the size of the stacked chain grows. Competing nuclear quantum effects make the net quantum contribution small, but very sensitive to anharmonicities, vdW, and number of HBs. Our results suggest that a reliable treatment of these systems can only be attained by considering all of these components.

Generally when dealing with complex systems like biomolecules, there is no easy answer as to which level of theory will be necessary to obtain a reliable structure or dynamics. The reason is that these systems are governed by a delicate balance of weak interactions, all of which can have equally large impacts on the final result. For example, taking into account the quantum nature not only of the electrons but also of the nuclei (here referred to as nuclear quantum effects, or NQE), tends to be important in systems stabilised by "weak" interactions like hydrogen bonds (HBs) and van der Waals (vdW) dispersion. NQE, in fact, can either act to stabilise or destabilise different structures.[1] The trends in stabilisation and destabilisation have been rationalised by the presence of competing NQE in inter- and intramolecular motions.[1,2] The effect of NQE on biological systems has been studied previously,[3,4] however the interplay of NQE with the HB "cooperativity effect"[5–17] and van der Waals interactions[17] in more complex biomolecular systems with degrees of freedom spanning very different anharmonic molecular motions has not yet been quantified. In the following we present a study, that goes beyond a standard study of H-bond cooperativity and role of vdW interactions, and also beyond the evaluation of NQE for simple H-bonded systems.

We take as a model stacked polyglutamine (polyQ) strands – a system that is relevant also to biological processes. Genetic mutations that lead to the expansion of polyQ sequences beyond a certain length in proteins are associated with many degenerative diseases, including Huntington disease.[6,18–21] It is proposed that these polyQ tracts favor the formation of $\beta$-strands which can then form $\beta$-hairpins and stack in aggregated H-bonded structures[6,18]. Before one can study the causes that lead to the stacking and aggregation of these structures in real peptides and biological environments, it is important to isolate and study the different qualitative contributions to their stabilisation. We here simply treat an isolated, infinitely periodic, antiparallel $\beta$-sheet model system (Fig. 1). Despite this simplification, the physics of the processes studied here should also be present in biological environments, even if embedded in more complex surrounding.

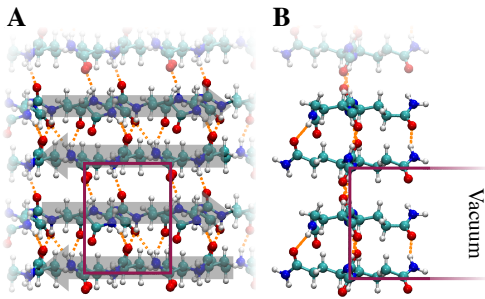

Figure 1: Schematic representation of (a) front view and (b) side view of the periodically repeating anti-parallel $\beta$-sheet polyglutamine (polyQ) structure. H-bonds are shown in orange. The red lines mark the unit cell. Carbon atoms are in turquoise, nitrogen atoms in blue, oxygen atoms in red, and hydrogen atoms in white.

The polyQ examined here are very flexible, presenting very anharmonic potential energy surfaces. In order to estimate free energies, not only the harmonic approximation should fail, but also empirical force fields which use harmonic terms for certain energetic contributions could yield unreliable results. In addition, it is known that for an accurate description of HB cooperativity, first-principles potential energy surfaces are necessary.[10,11,17] We thus opted to perform *ab initio* path integral molecular dynamics (PIMD) on a density-functional the-ory (DFT) potential energy surface to calculate dynamical and structural aspects, and the quantum contributions to binding free energies of these systems. In order to evaluate the free energy differences we perform mass thermodynamic integrations (mass-TI)[22–25] where we take the system to its classical limit by progressively increasing its mass. Overall we find vdW interactions and HB cooperativity to be of greater magnitude than NQE in stabilising the stacked structures. However, we find the contributions of NQE to be extremely sensitive to the proper treatment of the two effects above, and also to the inclusion or not of anharmonicities of the potential energy surface.

To begin, we fully relaxed a unit cell with two antiparallel polyQ strands using the FHI-aims program package[26] and the PBE[27]+vdW[28] functional. This structure is shown in Fig. 1. For more details of these calculations, see the SI. In order to calculate the build-up of cooperativity in this system we took the geometry obtained for the infinitely periodic structure and performed single point calculations starting from an isolated double-stranded structure, adding a new double strand at each step. These new structures are periodic only in the backbone direction and isolated in the other two, as shown pictorially in Fig. 2. By not performing further geometry optimisations we examine the stabilisation of the increasing stacked structures, approaching the infinite limit. We calculate

$$\Delta E_n = E(n) - E(n-2) - 2E_{opt}(1), \tag{1}$$

where $n=2, 4, 6, 8, 10$ is the number of polyglutamine strands considered, $E(n)$ is the total energy of the structure with $n$ strands, and $E_{opt}(1)$ is the total energy of the optimised single strand structure, here taken as our reference. In order to better visualize the build up of the cooperativity effect, we then divided each $\Delta E_n$ by the number of new HBs formed by the addition of an antiparallel double $\beta$-strand in that step. We show in Fig. 2 these quantities calculated with the PBE, PBE+vdW functional (pairwise van der Waals dispersion) and the PBE+MBD@rsSCS[29,30] functional, which takes into account many body contributions to dispersion up to infinite order. We observe a 100 meV/HB energy gain in binding strength when going from 2 to 4 stacked strands and then a slow non-linear increase of about 15 meV/HB in total until reaching the infinite limit. Adding vdW contributions is essential as it stabilises the structures by a factor of around 1.6, without affecting much the behavior of cooperativity (consistent with what was observed for peptide helices including or not vdW dispersion[17]). A similar cooperativity behavior for parallel isolated polyQ strands has been observed in Ref.,[8] albeit disregarding vdW contributions. Many body dispersion energy contributions destabilise the fully periodic structure by 16 meV per HB with respect to the pairwise vdW treatment. Though not negligible, it only represents a small fraction of the full vdW contributions.

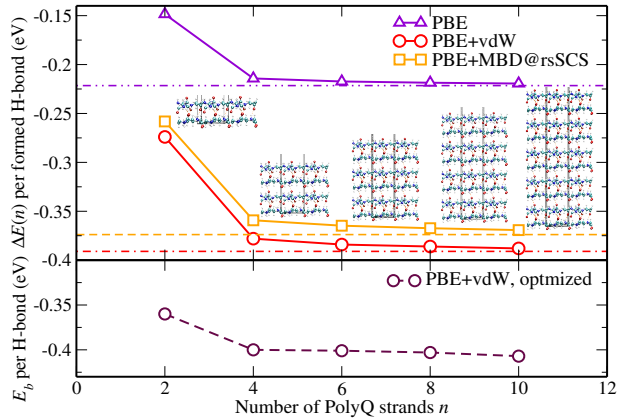

Figure 2: Top: Energy per added antiparallel double strand $\Delta E_n$, divided by the number of HBs formed in each step, calculated for the polyQ system. Curves are shown for the PBE[27] (purple triangles), PBE+vdW[28] (red circles) and PBE+MBD@rsSCS[29,30] (orange squares) functionals. Dotted and dashed lines represent the limit for the infinitely periodic system for the PBE, PBE+vdW, and PBE+MBD@rsSCS systems. Bottom: Binding energy of each system with the PBE+vdW functional, with all structures fully relaxed.

In order to assess the effect of relaxation on the structures, which is important in connection to the simulations involving dynamics presented below, we have fully relaxed all structures with the PBE+vdW functional. We evaluate a "binding" energy $E_b$ as

$$E_b = E_{opt}(n) - nE_{opt}(1), \qquad (2)$$

dividing this quantity by the total number of HBs at each $n$. This yields a curve also shown in Fig. 2. As expected the structures get stabilised with the relaxation, but the overall cooperativity trend can still be observed even if the energy drop from $n$=2 to 4 is now only around 40 meV. Having established that HB cooperativity and vdW dispersion play a critical role in stabilising polyQ structures and we turn to the role of NQE.

In order to determine the contributions of NQE to the binding free energy, we calculated the free energy difference in going from a classical to a quantum system by performing mass-TI. To this end we used the relation

$$\Delta F_{c \to q} = \int_{m_0}^{\infty} \frac{\langle K(\mu) - K_{class} \rangle}{\mu} d\mu = \\ = \int_0^1 \frac{2 \langle K(m_0/g^2) - K_{class} \rangle}{g} dg, \quad (3)$$

where $K$ is the kinetic energy, $m_0$ is the physical mass of the atoms, $\mu$ the mass integration variable, and $g = \sqrt{m_0/\mu}$. The brackets denote an ensemble average and the masses of all atoms are scaled in this procedure. This equation can be straightforwardly derived from the (quantum) partition function, and has been proposed in many forms in a number of previous studies.[22–24] Here we simply make use of the idea that the limit of infinite mass corresponds to the classical limit, where the quantum and the classical kinetic energy would be equal.

Table 1: $E_b$ as defined in Eq. 2, vibrational contribution to the "binding" free energy in the harmonic approximation $F_{vb}^{harm}$, and quantum contributions to the binding free energy in the harmonic approximation $(\Delta F)_b^{harm}$ and from PIMD calculations $(\Delta F)_b$ at 300K per total number of HBs (4 for $n$=2, and 10 for $n$=4). Energies in meV.

| $n$ | $E_b$ | $F_{vb}^{harm}$ | $(\Delta F)_b^{harm}$ | $(\Delta F)_b$ |
|---|---|---|---|---|
| 2 | -360 | 28 | 3 | -6 ± 2 |
| 4 | -400 | 29 | 3 | -3 ± 1.5 |

The quantity we are interested in calculating is the quantum contribution to the binding free energy $(\Delta F)_b$, here defined as

$$(\Delta F)_b = \Delta F_{c \to q}(n) - n \Delta F_{c \to q}(\text{strands}) =$$
$$= \int_0^1 \frac{2}{g} [\langle K_n(m_0/g^2) \rangle - n \langle K_{strands}(m_0/g^2) \rangle] dg, \quad (4)$$

where $\langle K_n \rangle$ is the average kinetic energy of the system with $n$ polyQ strands. Defined in this way, a negative $(\Delta F)_b$ means the quantum contributions have stabilised the H-bonded system.

We perform all calculations with the i-PI[31] wrapper code connected to the FHI-aims program package and the PBE+vdW functional. PBE+vdW has been shown to provide an accurate description for a broad range of hydrogen bonded and van der Waals bonded systems (see. e.g. Ref.[32]). Even though it is also known that generalised gradient approximations give a spurious softening of the H-bonds,[33,34] the computational cost of a hybrid functional would be prohibitive for the following calculations. Additional details about the calculations can be found in the SI. For comparison, we also calculated Eq. 4 in the harmonic approximation,[25] by calculating the phonons and harmonic vibrational density of states of the system.

We report in Table 1 $E_b$ as defined in Eq. 2, the vibrational free energies at 300K in the harmonic approximation, and the quantum contributions to the binding free energies in the harmonic approximation and from the path-integral mass integration per average total number of HBs present in the simulation. We see that in the harmonic approximation, zero point energy, temperature, and entropic contributions amount to around 10% of $E_b$, from which 10% is due to quantum contributions. However, while in the harmonic approximation NQE act to destabilise both the double and tetra-stranded quantum structures with respect to their classical counterparts, the anharmonic effects reverse these trends and predict a total stabilisation. Thus it is only with the full path integral treatment that the correct qualitative behaviour is observed.

The stabilisation effect is very similar for the double- and tetra-stranded structures –

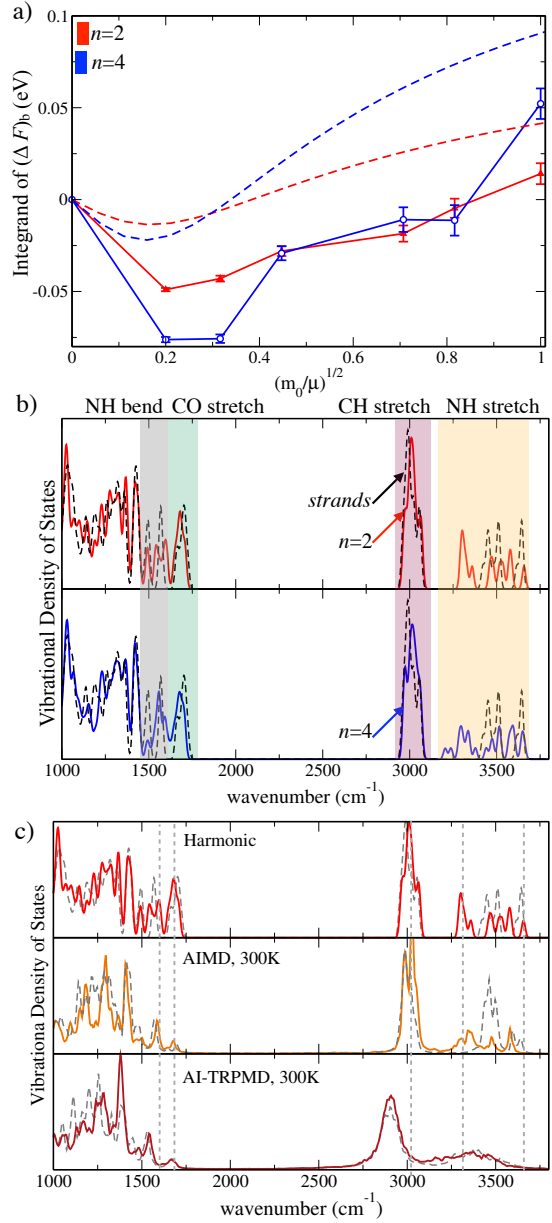

Figure 3: (a) Integrand of Eq. 4 in the harmonic approximation (dashed lines) and fully anharmonic (PIMD) case (full lines) for $n=2$ and $n=4$, on a PBE+vdW potential energy surface. (b) PBE+vdW harmonic vibrational density of states (VDOS) for the single polyQ strands (black dashed line), for the $n=2$ structure (red, upper pannel), and the $n=4$ structure (blue, lower pannel). (c) PBE+vdW VDOS in the harmonic approximation (upper panel), from the Fourier transform of the velocity autocorrelation function with classical nuclei at 300K (middle panel), and with quantum nuclei using the TRPMD method at 300K (lower panel) for $n=2$ and the respective single strands (shown as dashed lines). The simulations were run for 10 ps.

certainly within our errors, which are reported in Table 1. Also, even though the quantum contribution is small, we are here comparing quite different structures (isolated and H-bonded polyQ strands). Since for the H-bonded structures the total contributions are of the order of $k_bT$ at room temperature (see Fig. S2 of the SI), and in a more realistic situation there should be competing conformers within this energy window, these effects may actually play an important role.

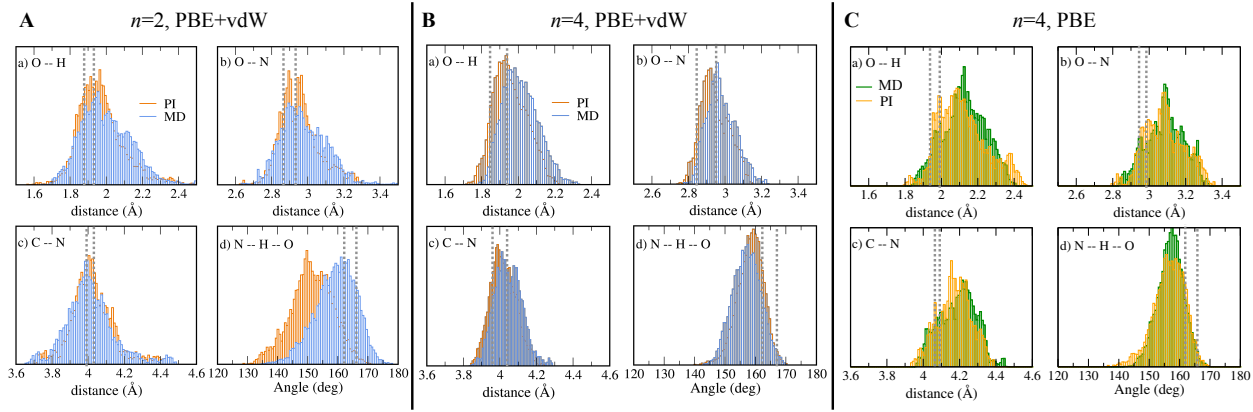

Figure 4: Structural quantities for the classical AIMD simulations and the PIMD ones, for A: $n=2$ with the PBE+vdW functional, B: $n=4$ with the PBE+vdW functional and C: $n=4$ with the PBE functional. We consider only backbone HBs and report $O\cdots H$ bond lengths, the $O\cdots N$ bond lengths, the $C\cdots N$ bond lengths (rough measure of the backbone distance), and the HB angles formed by $N\cdots H\cdots O$ atoms. The dotted lines in each plot correspond to the values for $n=2$ and the interval of values for $n=4$ for that distance or angle found in the respective fully relaxed structure.

In order to understand the origins of the observed NQE contributions, we plot the integrands of Eq. 4 in the harmonic approximation and from the path integral mass-TI, both shown in Fig. 3(a). The shapes are qualitatively similar, even if there is no *a priori* reason for them to exhibit the exact same shape, but anharmonic effects substantially pull the curves to lower energies. This suggests greater softening of the vibrational modes in the anharmonic case with respect to the harmonic one. Similar effects have been seen for low energy modes in other peptides including anharmonicities for classical nuclei.[35,36]

In Fig. 3(b) we show the changes in the vibrational density of states (VDOS) that we observe upon forming HBs, in the harmonic approximation. Especially in the region between 3200 and 3400 cm$^{-1}$ we observe a softening of the H-bonded NH stretches as $n$ increases, which is consistent with the strengthening of the HBs with chain length observed due to cooperativity. In the NH bending region it is hard to pinpoint differences since the spectra are very congested. It can be shown (see SI, Fig. S5) that the shape of the integrand in Fig. 3(a), at least in the harmonic approximation, does not come only from competing contributions (softening and hardening) from vibrational modes directly connected to the H-bonds, but from an intricate interplay involving destabilising contributions from backbone vibrations, CH stretches and NH bendings, and stabilising contributions from CO and NH stretches. In Fig. 3(c) we focus on the structure with 2 antiparallel polyQ strands and calculate as well the VDOS given by the Fourier transform of the velocity autocorrelation function from an *ab initio* MD (AIMD) run (i.e. with classical nuclei) and from the newly proposed thermostatted ring polymer molecular dynamics (TRPMD) method,[37] at 300K [1]. Comparing the AIMD case at 300 K with the harmonic one (0 K), we observe a softening of the high frequency NH stretches, but most of the rest of the spectrum remains the same. With TRPMD there is a much more pronounced softening of both NH and CH stretches by as much as 100 cm$^{-1}$. Since TRPMD produces typically broader peaks in the spectra,[37,38] it is hard to analyse exactly which modes are more softened upon H-bond formation in the quantum case (compare solid and dashed lines). However, the larger softening observed in the TRPMD VDOS can explain the pronounced dips in the integrand reported in 3(a) and the fact that the structures are stabilised by the quantum contributions in the fully anharmonic case. We stress that this effect would not be captured by classical anharmonicity alone.

Structural aspects can also give important physical insight. We performed PIMD and classical *ab initio* NVT MD simulations with the PBE and PBE+vdW functionals, for the double- and tetra-stranded structures for 10 ps, at 300K. We considered only backbone HBs and compared O···H bond lengths, O···N bond lengths, C···N bond lengths (rough measure of the backbone distance), and HB angles. These quantities are shown in Fig. 4A for $n$=2, in Fig. 4B for $n$=4 with the PBE+vdW functional, and in Fig. 4C for $n$=4 with the PBE functional. For $n$=2 the softening of the NH bendings is clearly reflected in the shift of the distribution of HB angles to lower values in the quantum case. It is hard to observe any other very pronounced effect except perhaps for a small shift to shorter O···H distances in the quantum case. Comparing to Fig. 3(c) middle and lower panels (AIMD and AI-TRPMD), this means that we would either need much more sampling to observe an effect here, or the backbone HBs (connected to the NH stretches) are not that much softened in the quantum case. When going to $n$=4 with the PBE+vdW functional, the effects are more pronounced. All distances are shifted to smaller values in the quantum case – even if just barely for the backbone. The differences in angles between the classical and quantum cases is small, but now the angles are more stiff in the quantum case (this stiffening trend of the angles is consistent with Ref.[1] for stronger HBs). This points to a considerable hardening due to NQE of the NH bendings in the tetra-stranded structure, which is again consistent with the competing NQE picture and could explain why we do not see a much larger energetic stabilisation for the $n$=4 than for the $n$=2 case. It is possible that for even larger structures the stabilisation brought about by the softening of the NH stretches wins over. It is interesting to note in Fig. 4C that when analysing the dynamics of $n$=4 with the PBE functional both the distances between atoms and the angle distributions are essentially on top of one another even if the distances show considerably larger mean values than in the PBE+vdW case. This is to be expected, since these H-bonds are weaker than the PBE+vdW ones.

---

[1]TRPMD is equivalent to PIMD for static properties, but in addition can also give an approximation to time correlation functions.

In passing, we mention that we also performed mass-TI simulations changing only the masses of the hydrogen atoms in the system to reach that of deuterium, for the isolated strands and for $n$=2. We found that the deuterated structure is ever so slightly destabilised, but only by 1 meV per HB. Within our error bars this value is not significant but perhaps worth reporting.

We have reported a systematic study on the combination of so-called "weak" interactions that contribute to structure stabilisation of model antiparallel stacked polyQ (a peptide repeat often found in amyloid diseases), namely, HB cooperativity, vdW dispersion interactions, and NQE including dynamics and a fully anharmonic potential energy surface. We have drawn a complex picture of the interplay of these effects in larger biomolecular systems that had not been previously quantified.

We find that among these interactions, vdW dispersion represents the largest contribution to H-bond stabilisation, followed by HB cooperativity as the chain grows. The impact of NQE, though small (a few meV per H-bond), is seen to depend on its interplay with cooperativity, vdW, and whether one takes into account the full anharmonicity of the potential energy surface or not. Anharmonic effects change the sign of the NQE contributions to the free energy with respect to the harmonic estimate, most likely due to the observed overall softening of the vibrational modes caused by the inclusion of NQE on dynamics. We thus find that in an anharmonic picture, NQE represent a small stabilisation component in these structures. The strengthening of the HBs by vdW and cooperativity also enhances the competing nature of NQEs. Structural properties are visibly affected by NQE differently depending on the size of the chain and the addition of vdW, but for the energy contributions we calculated, we did not observe a strong impact, as long as one goes beyond the harmonic approximation. These effects may be important for structures that are in close energetic competition, e.g. slightly different folding motifs.

The system studied here is just a first step in order to understand these structures from a quantum mechanical perspective. With growing computational capacity, it would be ideal to employ better electronic structure methods (NQE on structural properties can be affected by the choice of potential) and consider systems closer to the biological reality. Nevertheless, the results presented here suggest that for a reliable prediction of stability of close-competing conformational motifs in larger polypeptides – which can be decisive in studying folding and misfolding of proteins – a high level (quantum) treatment of nuclei, electronic structure, and anharmonicities is necessary. This, of course, also underlines the need to train empirical potentials on high-level potential energy surfaces.

## Acknowledgement

The authors thank Michele Ceriotti and David Manolopoulos for relevant discussions and input in the paper. M.R. acknowledges funding from the German Research Foundation (DFG) under project RO 4637/1-1, A.M. is supported by the European Research Council under the European Union's Seventh Framework Programme (FP/2007-2013) / ERC Grant Agreement number 616121 (HeteroIce project) and the Royal Society through a Royal Society Wolfson Research Merit Award. We are grateful to the UKCP consortium (EP/ F036884/1) for access to ARCHER.

## Supporting Information Available

Supplemental information about technical details of the simulations and convergence tests are available.

This material is available free of charge via the Internet at `http://pubs.acs.org/`.

# Graphical TOC Entry

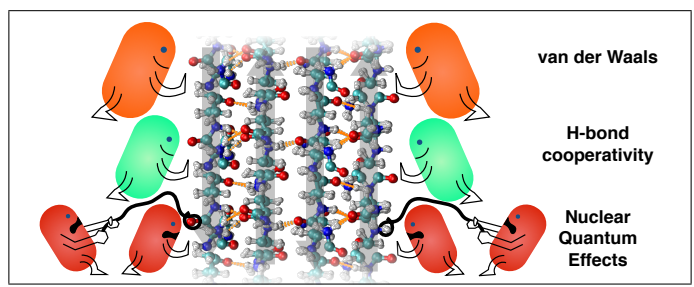

# Supplemental Information: Stability of Complex Biomolecular Structures: Van der Waals, Hydrogen Bond Cooperativity, and Nuclear Quantum Effects

Mariana Rossi*
*Physical and Theoretical Chemistry Lab, University of Oxford, South Parks Road, OX1 3QZ Oxford, UK†*

Wei Fang and Angelos Michaelides
*Thomas Young Centre, London Centre for Nanotechnology,
and Department of Chemistry, University College London,
17-19 Gordon Street, WC1H 0AH London, UK*

## I. COMPUTATIONAL DETAILS AND CONVERGENCE TESTS

The initial structure for a single $\beta$-strand of polyglutamine was generated with the TINKER[1] program package, setting the backbone angles consistent with the ideal conformation for anti-parallel $\beta$-sheets. We chose the antiparallel conformation since this seems to be the most stable structure adopted by polyQ repeats in peptides[2]. The initial structure for the anti-parallel $\beta$-sheet was then obtained by aligning two strands in a 'head-to-toe' arrangement. Our unit cell for this $\beta$-sheet contains two glutamine peptides on the bottom strand and two on the upper strand (68 atoms). We aligned the backbone along the $x$ direction (H-bonds roughly along the $z$-axis) and calculated the lattice parameters that would create an infinitely repeating ideal structure along $x$ and $z$. We then fully relaxed the periodic structure and the unit cell, insulating it with a 100 Å vacuum in the $y$ direction, using the FHI-aims program package[3], DFT with the PBE[4]+vdW[5] functional, *tight* basis set and numeric integration settings, and a k-point grid of $6\times1\times4$ (which proved to be already beyond convergence since the system is strongly isolating). The final structure had lattice constants $a$=6.971Å, $b$=100Å, $c$=9.070Å, $\alpha$=90.0$^o$, $\beta$=88.5$^o$, and $\gamma$=90$^o$. We identified six HBs in the unit cell, using a simple distance criterion ($d < 2.2$ Å): four coming from the backbone and two coming from the side chains of glutamine.

We evaluated $\langle K \rangle$ with the i-PI[6] wrapper code connected to FHI-aims, where density functional theory calculations were performed with the PBE+vdW functional, a $5\times1\times1$ k-point grid and *light* settings for the basis sets and integration grids. We performed the PIMD simulations using the PIGLET thermostat[7] at a temperature of 300K, using 6 replicas of the system, an integration time step of 0.5 fs (and 1 fs for the higher masses) and six points for the mass integration, corresponding to factors of 1, 1.5, 2, 5, 10, and 25 times the nominal mass. We calculated the kinetic centroid virial estimator for the kinetic energy for each mass point. For the smaller mass points we performed 2 ps of thermalization plus at least 8 ps of simulation for each integration point (for the isolated strand and for the structures with 2 and 4 strands). For the higher masses we could use a 1 fs time step. Convergence tests are shown in Fig. S2. We chose to use the mass-TI instead of scaled coordinate method[8] because we can perform path integral simulations using the PIGLET[7] colored noise thermostat in order to reduce the number of beads of the path integral ring polymer. The change of variables in the integrand of Eq. 4 of the manuscript from $\mu$ to $g$ makes the integrand flatter for lower masses, where nuclear quantum effects are more pronounced – thus requiring fewer integration points (see Fig. S1).

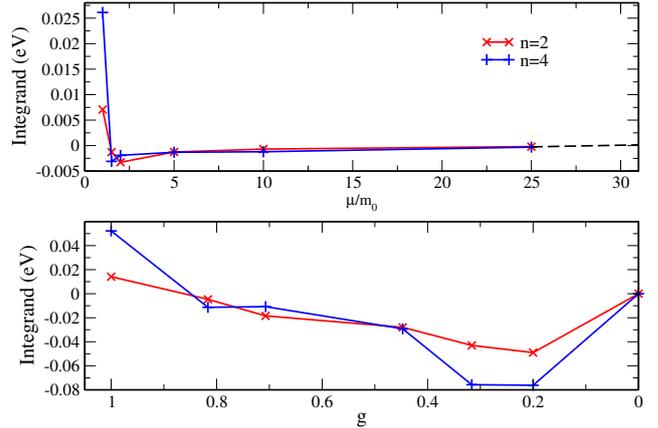

FIG. S1. Comparison of the integrand that goes into the mass integration expression for $(\Delta F)_b$ when using or not using the change of variables $g = \sqrt{m_0/\mu}$.

For the largest system (4 strands, 136 atoms) each force evaluation (per replica) took 56 seconds parallelised over 96 processors in the ARCHER supercomputer (Cray XC30) – and each mass integration point consisted of 6 replicas and between 8000 and 20000 force evaluations.

For the evaluation of phonons, we used the phonopy program package[9] and FHI-aims[3], thus obtaining phonons through finite differences (we used 2 repetitions of the initial unit cell). We calculated $(\Delta F)_b^{harm}$ analytically and through the harmonic mass-TI using the phonon density of states.

In order to evaluate the vibrational densities of states (VDOS) in Fig. 3c of the manuscript, we ran 10 ps of PBE+vdW AIMD-NVE simulations and AI-TRPMD with 16 replicas of the system and a 0.5 fs time step, starting from previous thermalizations at 300K. In Figure S3 we plot, as an example, the VDOS as obtained after 5 ps of simulation and after 10 ps for $n=2$. Except for changes in relative heights of the peaks, the peak positions, which is what we rely on for the discussion in the manuscript, are expected to be converged at 10 ps. In Figure S4 we show a zoom in the low frequency range of the vibrational density of states coming from AIMD (classical nuclei) simulations (orange), and TRPMD simulations (red) for $n=1, 2$. Even if the region is very congested we observe more pronounced shifts between the orange and the red lines for peaks lying above 1000 cm$^{-1}$.

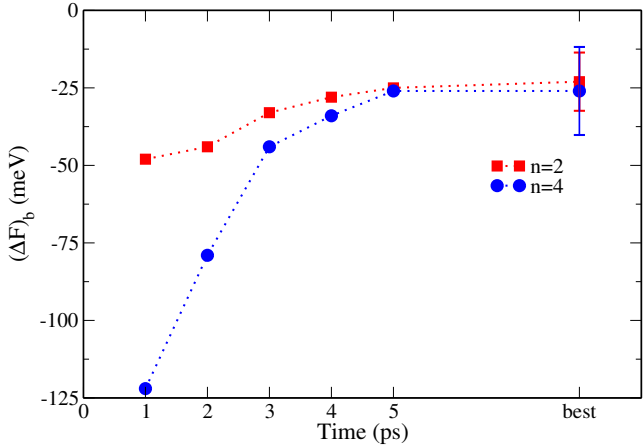

FIG. S2. Convergence of total anharmonic $(\Delta F)_b$ at 300K (PBE+vdW) with time of simulation for $n=2$ and 4. Error bars shown only for the converged point.

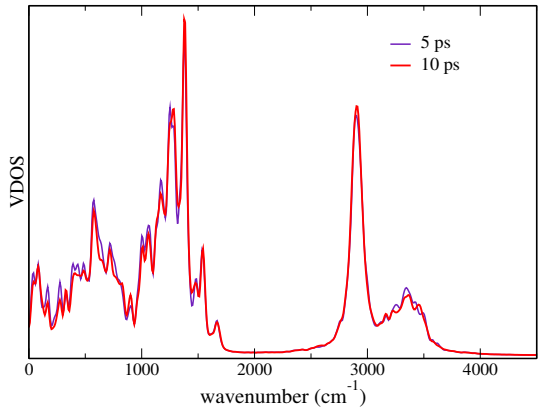

FIG. S3. Vibrational density of states obtained with TRPMD for $n=2$ as obtained after 5 ps and after 10 ps of simulation.

## II. DECOMPOSITION OF COMPETING NUCLEAR QUANTUM EFFECTS

In Figure S5 we show a decomposition of the integrand of Equation 4 of the manuscript in the *harmonic* approximation plotting separately the quantum contribution to the free energy coming from frequencies above and be-low a certain threshold $f$. The shape of this integrand is also shown in Fig. 3a of the manuscript. When decomposing it in contributions coming from frequencies above and below $f$ it is possible to analyze which modes are contributing for a stabilization and which are contributing for a destabilization In Fig. S5A we show the integral (quantum contribution to the free energy in the harmonic approximation) for several values of $f$ considering frequencies $\omega$ greater than $f$ and lesser than $f$. We also show in shaded grey the regions where backbone CO bending (CO-b), NH bending (NH-b), CO stretches (CH-s) and NH stretches (NH-s) are observed from the normal mode analysis. For specific positions, which are color coded in panel A, we show also the shape of the integrand in Fig. S5B, also separating it in what comes from $\omega > f$ and $\omega < f$.

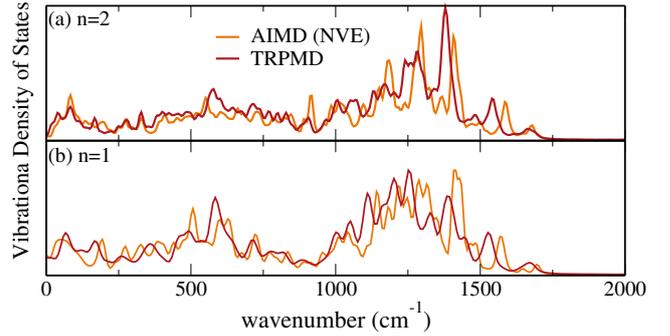

FIG. S4. Low frequency range of the vibrational density of states for $n=2$ (top) and $n=1$ (bottom) as obtained with the PBE+vdW functional in a TRPMD simulation (red curves) and a AIMD simulation (orange curves).

In Fig. S5A while the orange crosses show a decreasing behavior (black plus signs show increasing behavior) we are adding to the black curve modes that contribute to destabilization and excluding them from the orange curve. When it flattens (e.g. between 850 and 1050 cm$^{-1}$) the frequencies in that range do not contribute to stabilization or destabilization. Finally when there is an inverted behavior, e.g. around 1600 and 1700 cm$^{-1}$, we are including in the black curve vibrations that stabilize the structure and excluding them from the orange curve. One thing to note is that the backbone CO bends are in a frequency range that would be almost fully classical at room temperature since $k_B T$ at 300K corresponds to 208 cm$^{-1}$ and do not contribute to stabilization or destabilization. Another point to note is that the competing effect is of a very intricate nature, with modes not directly connected to the H-bonds also contributing to the picture. As expected, both the CO stretches and the NH stretches contribute to an stabilization, while the destabilization comes from the NH bendings and also other bending motions involving the backbone and and side-chain groups, as well as CH stretches (around 3000 cm$^{-1}$). Since when including anharmonicities the coupling between modes increases, we suppose that the com-peting nature of the nuclear quantum effects in this system cannot always be rationalised solely in terms of the H-bond distance.

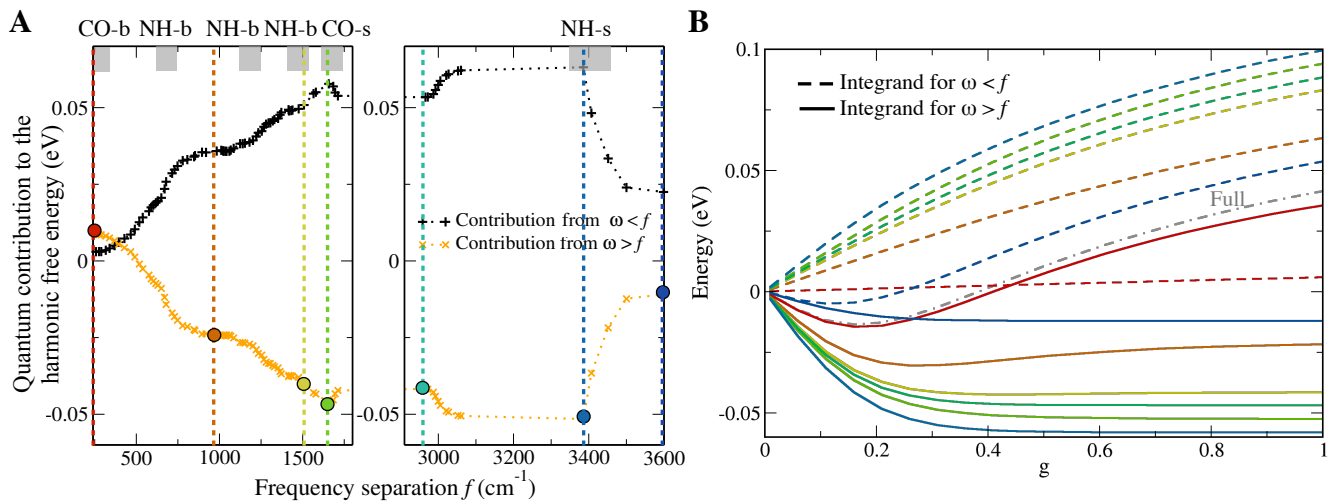

FIG. S5. A: Quantum contribution to the free energy in the harmonic approximation coming from frequencies $\omega$ above $f$ (orange crosses) and below $f$ (plus signs). Approximate regions corresponding to backbone CO bending (CO-b), NH bending (NH-b), CO stretches (CH-s) and NH stretches (NH-s) in the normal mode analysis are shaded in grey in the top of the plot. The addition of the orange and black curves at each point yields the full contribution to the total energy. B: For color coded positions (dashed lines and circles) in panel A we show the shape of the integrand of Eq. 4 in the manuscript also separated in contributions coming from $\omega > f$ (solid lines) and from $\omega < f$ (dashed lines).